



\hsize=6.0truein
\vsize=8.5truein
\voffset=0.25truein
\hoffset=0.1875truein
\tolerance=1000
\hyphenpenalty=500
\def\monthintext{\ifcase\month\or January\or February\or
   March\or April\or May\or June\or July\or August\or
   September\or October\or November\or December\fi}


\font\tenrm=cmr10 scaled \magstep1   \font\tenbf=cmbx10 scaled \magstep1
\font\sevenrm=cmr7 scaled \magstep1  
\font\fiverm=cmr5 scaled \magstep1   

\font\teni=cmmi10 scaled \magstep1   \font\tensy=cmsy10 scaled \magstep1
\font\seveni=cmmi7 scaled \magstep1  \font\sevensy=cmsy7 scaled \magstep1
\font\fivei=cmmi5 scaled \magstep1   \font\fivesy=cmsy5 scaled \magstep1

\font\tentt=cmtt10 scaled \magstep1
\font\tenit=cmti10 scaled \magstep1
\font\tensl=cmsl10 scaled \magstep1

\def\twelvepoint{\def\rm{\fam0\tenrm}
   \textfont0=\tenrm \scriptfont0=\sevenrm \scriptscriptfont0=\fiverm
   \textfont1=\teni  \scriptfont1=\seveni  \scriptscriptfont1=\fivei
   \textfont2=\tensy \scriptfont2=\sevensy \scriptscriptfont2=\fivesy
   \textfont\itfam=\tenit \def\it{\fam\itfam\tenit}
   \textfont\ttfam=\tentt \def\tt{\fam\ttfam\tentt}
   \textfont\bffam=\tenbf \def\bf{\fam\bffam\tenbf}
   \textfont\slfam=\tensl \def\sl{\fam\slfam\tensl} \rm
   \hfuzz=1pt\vfuzz=1pt
   \setbox\strutbox=\hbox{\vrule height 10.2pt depth 4.2pt width 0pt}
   \parindent=24pt\parskip=1.2pt plus 1.2pt
   \topskip=12pt\maxdepth=4.8pt\jot=3.6pt
   \normalbaselineskip=14.4pt\normallineskip=1.2pt
   \normallineskiplimit=0pt\normalbaselines
   \abovedisplayskip=13pt plus 3.6pt minus 5.8pt
   \belowdisplayskip=13pt plus 3.6pt minus 5.8pt
   \abovedisplayshortskip=-1.4pt plus 3.6pt
   \belowdisplayshortskip=13pt plus 3.6pt minus 3.6pt
   \topskip=12pt \splittopskip=12pt
   \scriptspace=0.6pt\nulldelimiterspace=1.44pt\delimitershortfall=6pt
   \thinmuskip=3.6mu\medmuskip=3.6mu plus 1.2mu minus 1.2mu
   \thickmuskip=4mu plus 2mu minus 1mu
   \smallskipamount=3.6pt plus 1.2pt minus 1.2pt
   \medskipamount=7.2pt plus 2.4pt minus 2.4pt
   \bigskipamount=14.4pt plus 4.8pt minus 4.8pt}

\twelvepoint



\font\titlerm=cmr10 scaled \magstep3
\font\titlerms=cmr10 scaled \magstep1 
\font\titlei=cmmi10 scaled \magstep3  
\font\titleis=cmmi10 scaled \magstep1 
\font\titlesy=cmsy10 scaled \magstep3 	
\font\titlesys=cmsy10 scaled \magstep1  
\font\titleit=cmti10 scaled \magstep3	
\skewchar\titlei='177 \skewchar\titleis='177 
\skewchar\titlesy='60 \skewchar\titlesys='60 

\def\titlefont{\def\rm{\fam0\titlerm}
   \textfont0=\titlerm \scriptfont0=\titlerms 
   \textfont1=\titlei  \scriptfont1=\titleis  
   \textfont2=\titlesy \scriptfont2=\titlesys 
   \textfont\itfam=\titleit \def\it{\fam\itfam\titleit} \rm}


\def\preprint#1{\baselineskip=19pt plus 0.2pt minus 0.2pt \pageno=0
   \begingroup
   \nopagenumbers\parindent=0pt\baselineskip=14.4pt\rightline{#1}}
\def\title#1{
   \vskip 0.9in plus 0.45in
   \centerline{\titlefont #1}}
\def\secondtitle#1{}
\def\author#1#2#3{\vskip 0.9in plus 0.45in
   \centerline{{\bf #1}\myfoot{#2}{#3}}\vskip 0.12in plus 0.02in}
\def\secondauthor#1#2#3{}
\def\addressline#1{\centerline{#1}}
\def\abstract{\vskip 0.7in plus 0.35in
	\centerline{\bf Abstract}
	\smallskip}
\def\finishtitlepage#1{\vskip 0.8in plus 0.4in
   \leftline{#1}\supereject\endgroup}

\def\date#1{\finishtitlepage{#1}}

\def\nolabels{\def\eqnlabel##1{}\def\eqlabel##1{}\def\figlabel##1{}%
	\def\reflabel##1{}}
\def\writelabels{\def\eqnlabel##1{%
	{\escapechar=` \hfill\rlap{\hskip.11in\string##1}}}%
	\def\eqlabel##1{{\escapechar=` \rlap{\hskip.11in\string##1}}}%
	\def\figlabel##1{\noexpand\llap{\string\string\string##1\hskip.66in}}%
	\def\reflabel##1{\noexpand\llap{\string\string\string##1\hskip.37in}}}
\nolabels


\global\newcount\secno \global\secno=0
\global\newcount\meqno \global\meqno=1
\global\newcount\subsecno \global\subsecno=0

\font\secfont=cmbx12 scaled\magstep1

\def\section#1{\global\advance\secno by1
   \xdef\secsym{\the\secno.}
   \global\subsecno=0
   \global\meqno=1\bigbreak\medskip
   \noindent{\secfont\the\secno. #1}\par\nobreak\smallskip\nobreak\noindent}

\def\subsection#1{\global\advance\subsecno by1
\medskip
\noindent
{\bf\the\secno.\the\subsecno\ #1}
\par\medskip\nobreak\noindent}

\def\newsec#1{\global\advance\secno by1
   \xdef\secsym{\the\secno.}
   \global\meqno=1\bigbreak\medskip
   \noindent{\bf\the\secno. #1}\par\nobreak\smallskip\nobreak\noindent}
\xdef\secsym{}

\def\appendix#1#2{\global\meqno=1\xdef\secsym{\hbox{#1.}}\bigbreak\medskip
\noindent{\bf Appendix #1. #2}\par\nobreak\smallskip\nobreak\noindent}


\def\eqnn#1{\xdef #1{(\secsym\the\meqno)}%
	\global\advance\meqno by1\eqnlabel#1}
\def\eqna#1{\xdef #1##1{\hbox{$(\secsym\the\meqno##1)$}}%
	\global\advance\meqno by1\eqnlabel{#1$\{\}$}}
\def\eqn#1#2{\xdef #1{(\secsym\the\meqno)}\global\advance\meqno by1%
	$$#2\eqno#1\eqlabel#1$$}


\font\footfont=cmr10 scaled\magstep0

\def\shoe#1#2{{\footfont\raggedbottom\noindent\baselineskip=10pt plus
0.3pt\footnote{#1}{#2}}}

\def\myfoot#1#2{{\baselineskip=14.4pt plus 0.3pt\footnote{#1}{#2}}}
\global\newcount\ftno \global\ftno=1
\def\foot#1{{\baselineskip=14.4pt plus 0.3pt\footnote{$^{\the\ftno}$}{#1}}%
	\global\advance\ftno by1}


\global\newcount\refno \global\refno=1
\newwrite\rfile

\def\ref{[\the\refno]\nref}
\def\nref#1{\xdef#1{[\the\refno]}\ifnum\refno=1\immediate
	\openout\rfile=refs.tmp\fi\global\advance\refno by1\chardef\wfile=\rfile
	\immediate\write\rfile{\noexpand\item{#1\ }\reflabel{#1}\pctsign}\findarg}
\def\findarg#1#{\begingroup\obeylines\newlinechar=`\^^M\passarg}
	{\obeylines\gdef\passarg#1{\writeline\relax #1^^M\hbox{}^^M}%
	\gdef\writeline#1^^M{\expandafter\toks0\expandafter{\striprelax #1}%
	\edef\next{\the\toks0}\ifx\next\null\let\next=\endgroup\else\ifx\next\empty%

\else\immediate\write\wfile{\the\toks0}\fi\let\next=\writeline\fi\next\relax}}
	{\catcode`\%=12\xdef\pctsign{
\def\striprelax#1{}

\def\semi{;\hfil\break}
\def\addref#1{\immediate\write\rfile{\noexpand\item{}#1}} 

\def\listrefs{\vfill\eject\immediate\closeout\rfile
   {{\secfont References}}\bigskip{\frenchspacing%
   \catcode`\@=11\escapechar=` %
   \input refs.tmp\vfill\eject}\nonfrenchspacing}

\def\startrefs#1{\immediate\openout\rfile=refs.tmp\refno=#1}


\global\newcount\figno \global\figno=1
\newwrite\ffile
\def\fig{\the\figno\nfig}
\def\nfig#1{\xdef#1{\the\figno}\ifnum\figno=1\immediate
	\openout\ffile=figs.tmp\fi\global\advance\figno by1\chardef\wfile=\ffile
	\immediate\write\ffile{\medskip\noexpand\item{Fig.\ #1:\ }%
	\figlabel{#1}\pctsign}\findarg}

\def\listfigs{\vfill\eject\immediate\closeout\ffile{\parindent48pt
	\baselineskip16.8pt{{\secfont Figure Captions}}\medskip
	\escapechar=` \input figs.tmp\vfill\eject}}

\def\noblackbox{\overfullrule=0pt}
\def\inv{^{\raise.18ex\hbox{${\scriptscriptstyle -}$}\kern-.06em 1}}
\def\dup{^{\vphantom{1}}}
\def\Dsl{\,\raise.18ex\hbox{/}\mkern-16.2mu D} 
\def\dsl{\raise.18ex\hbox{/}\kern-.68em\partial}
\def\slash#1{\raise.18ex\hbox{/}\kern-.68em #1}
\def\lspace{}
\def\lbspace{}
\def\boxeqn#1{\vcenter{\vbox{\hrule\hbox{\vrule\kern3.6pt\vbox{\kern3.6pt
	\hbox{${\displaystyle #1}$}\kern3.6pt}\kern3.6pt\vrule}\hrule}}}
\def\mbox#1#2{\vcenter{\hrule \hbox{\vrule height#2.4in
	\kern#1.2in \vrule} \hrule}}  
\def\bar{\overline}
\def\e#1{{\rm e}^{\textstyle#1}}
\def\del{\partial}
\def\curly#1{{\hbox{{$\cal #1$}}}}
\def\curlyD{\hbox{{$\cal D$}}}
\def\curlyL{\hbox{{$\cal L$}}}
\def\vev#1{\langle #1 \rangle}
\def\psibar{\overline\psi}
\def\lform{\hbox{$\sqcup$}\llap{\hbox{$\sqcap$}}}
\def\darr#1{\raise1.8ex\hbox{$\leftrightarrow$}\mkern-19.8mu #1}
\def\half{{\textstyle{1\over2}}} 
\def\roughly#1{\ \lower1.5ex\hbox{$\sim$}\mkern-22.8mu #1\,}
\def\MSbar{$\bar{{\rm MS}}$}
\hyphenation{di-men-sion di-men-sion-al di-men-sion-al-ly}

\parindent=0pt
\parskip=5pt

\def\l{\lambda}
\def\r{\rho}
\def\w{\omega}
\def\t{\tau}
\def\tP{\tilde P}
\def\B{{\cal B}}
\def\P{\cal P}
\def\s{\sigma}
\def\k{\kappa}
\def\a{\alpha}
\def\b{\beta}
\def\al{\alpha_{(i)}}
\def\bigsum{\sum_{l=1}^{q-1}\sum_{k=1}^\infty}
\def\D{{\cal D}}
\def\O{\cal O}
\def\Donetwo{\D^{ij}_{1,2}}
\def\TP{\tilde P}
\def\Done{\D^{ij}_1}
\def\Dtwo{\D^{ij}_2}
\def\R{{\cal R}}
\def\Rk{{\cal R}_k}
\def\Rs{\hat{\rm R}}
\def\tk{t_k}
\def\Rm{{\cal R}_m}
\def\tm{t_m}
\def\Ok{{\cal O}_k}
\def\Om{{\cal O}_m}
\def\O#1{{\cal O}_{#1}}
\def\t#1{t_{#1}}
\def\Rlk{{\cal R}_{l,k}}
\def\Rjlk{{\cal R}^j_{l,k}}
\def\Rilk{{\cal R}^i_{l,k}}
\def\tlk{ t_{l,k}}
\def\rline{{{\rm I}\!{\rm R}}}
\def\pq{[P,Q]=1}
\def\pqq{[{\tilde P},Q]=Q}
\def\zplus{z\rightarrow +\infty}
\def\zminus{z\rightarrow -\infty}
\def\hf{\half}
\def\nonp{non-perturbative}
\def\hmm{hermitian matrix model}
\def\integ#1#2#3{\int_{#1}^{#2}\!\! d#3\ }
\def\TO{\tau_{\rm open}}
\def\TC{\tau_{\rm closed}}


\preprint{
\vbox{
\rightline{IASSNS--HEP--93/5}
\vskip2pt\rightline{hep-th/9301112}
\vskip2pt\rightline{January, 1993.}
}
}
\vskip -1cm

\title{On Integrable $c<1$ Open--Closed String Theory}
\author{\bf Clifford V. Johnson\myfoot{$^*$}{\rm Lindemann
Fellow}\myfoot{$^\dagger$}{\rm e-mail: cvj@guinness.ias.edu}}{}{}
\vskip 1.5cm
\addressline{\it School of Natural Sciences}
\addressline{\it Institute for Advanced Study}
\addressline{\it Olden Lane}
\addressline{\it Princeton, NJ 08540 U.S.A.}
\vskip -1cm

\abstract
The integrable structure of open--closed string theories in the $(p,q)$
conformal minimal model
backgrounds is presented. The relation between the $\tau$--function of the
closed string theory and that of the open--closed string theory is uncovered.
The
resulting description of the open--closed string theory is shown to fit very
naturally
into the framework of the $sl(q,{\rm C})$ KdV hierarchies. In particular, the
twisted
bosons which underlie and organise the  structure of the closed string theory
play a similar role here and may be employed to derive loop equations and
correlation function recursion relations for the open--closed strings in a
simple way.

\vskip -1cm
\date{}

\def\nuke{Nucl.Phys.}
\def\pl{Phys.Lett.}
 \nref\three{
E.Br\'{e}zin and V.Kazakov, Phys.Lett. {\bf B236} (1990) 144\semi
M.Douglas and S.H.Shenker, Nucl.Phys. {\bf B335} (1990) 635\semi
D.J.Gross and A.A.Migdal, Phys.Rev.Lett. {\bf 64} (1990) 127\semi
D.J.Gross and A.A.Migdal, Nucl. Phys. {\bf B}340 (1990) 333.}
\nref\douglas{M.Douglas, Phys.Lett. {\bf B238} (1990) 176.}
\nref\Gelfand{I.M.Gel'fand and L.A.Dikii, Russian Math. Surveys {\bf 30:5}
(1975) 77.}

\section{Introduction}
Much progress has been achieved in the understanding of  $c\leq 1$ string
theory using matrix models. In particular the structure of  closed string
theory
propagating in the  $(p,q)$ conformal minimal model backgrounds (with diagonal
modular invariants)  has been
studied
extensively. These $c<1$ string theories  have an underlying  integrable
hierarchy
structure which is that of the $(q-1)$th generalised KdV hierarchies associated
to  the Lie algebra $sl(q,{\rm C})$. A double scaled $(q-1)$--hermitian matrix
model
 is believed to contain (at least) this generalised KdV organisation of the
operator
content.
Alternatively, an appropriately tuned two--matrix model has been shown to
realise these models\ref\twomat{M.Douglas, {\sl `The Two--matrix Model'}, in
the proceedings of the Cargese Workshop {`Random Surfaces and Quantum
Gravity'}, May 27 to June 2, 1990. Plenum Press, New York 1991\semi
T.Tada and M.Yamaguchi, Phys. Lett. {\bf B 250} (1991) 38\semi
T.Tada, Phys. Lett. {\bf B259} (1991) 442\semi
J.M.Daul, V.Kazakov and I.Kostov, CERN preprint CERN--TH.6834/93, hepth
9303093.}.

The simplest example is the one--hermitian matrix model
\eqn\matrixmodel{Z=\int\!\!dM\exp{\left\{-{N\over\lambda}
{\rm Tr}[V(M)]
\right\}}}
which describes Euclidean 2D quantum gravity coupled to the $(2m-1,2)$
conformal
minimal models\three\ to all orders in genus perturbation
theory\shoe{$^\dagger$}{All
of the results presented in this paper will be taken to be valid only to all
orders in  the world--sheet expansion.} in the appropriate continuum limit.
This continuum limit is the `double scaling limit', where the size $N$ of the
hermitian matrix $M$ is sent to infinity while  ratios of couplings in the
polynomial matrix model potential $V(M)$ are tuned to finite values. These 2D
quantum gravity theories may be also viewed as closed string theories
`propagating' in the $(2m-1,2)$ conformal minimal model backgrounds. These
theories have an underlying integrable hierarchy structure which arises as
follows:
 The background fields of the string theory are the $ sl(2,{\rm C})$ KdV times
$\tk$, coupling to
scaling operators $\Ok$ living on the world sheet. The KdV equations organise
the operator structure and are usually written in terms of the `string
susceptibility' $u$ as follows:
\eqn\kdvflow{{\partial u(\tk)\over\partial\tk} =\R^{'}_{k+1}[u(\tk)],}
where the $\Rk$ are the Gel'fand--Dikii differential polynomials and a prime
denotes a differentiation with respect to $x\propto t_0$. In the unitary model
in this series, $x$ (the scaling part of $\lambda$) is the cosmological
constant of the theory and $u$ is the
two--point function of the `puncture operator' $\P$ of the theory which couples
to $x$. The KdV equations are supplied with `initial conditions' by the string
equation, which shall be written as \eqn\stringeq{\R=0,} where
\eqn\defr{\R\equiv\sum_{k=0}^\infty(k+\hf)\tk\Rk.} This information is enough
to fully determine the partition function of the closed string theory.

With the success of this formulation in mind, it is natural to study the
theories representing interacting open and closed strings in these backgrounds.
An obvious question to ask is whether the integrable structures underlying the
closed string  may be extended to this more general theory. This paper will
answer this question in the affirmative.
In ref.\ref\kostovii{I.K.Kostov, \pl {\bf B238}, (1990) 181.} the double
scaling limit of the above
one--hermitian matrix model  supplemented with a logarithmic potential
\eqn\logpot{{\gamma\over N}{\rm Tr}\log{(1-\mu^2M^2)}} was studied.  This
logarithm has the effect of adding surfaces with
boundaries of finite extent to the partition sum. The resulting models  are the
natural extensions of the above, representing the theories defined on all
smooth orientable 2D topologies. The string equation was derived
as:\eqn\openone{\R+2\Gamma\nu\Rs(x,\rho)=0,} where $\Gamma$, the scaling part
of $\gamma$, is the ratio of the
open--closed string coupling and the closed string coupling $\nu$ (renormalised
$1/N$). The parameter $\rho$, which is the scaling part of $\mu^2$,
is the  mass of the ends of the open strings. $\Rs$ is the diagonal part of the
resolvent of the Hamiltonian ${\cal H}\equiv-\nu^2\partial^2_x+u(x,\tk)$ which
satisfies the Gel'fand--Dikii
equation\Gelfand:\eqn\geldik{4(u+\rho)\Rs^2-2\Rs\Rs^{''}+(\Rs^{'})^2=1.} This
is the theory of open--closed strings in  the $(2m-1,2)$ background with the
naive
embedding dimension being  a single point. The closed string theory is
trivially recovered in the $\Gamma\to 0$ limit. The bulk operator content is
the same as
that of the closed string theory, possessing the KdV organisation shown above,
etc. In this paper it will be shown how the boundary operator content of the
closed string theory is related to that of the more general theory.

In ref.\ref\unitary{S.Dalley, C.V.Johnson, T.R.Morris and A.W\"atterstam, Mod.
Phys. Lett. {\bf A7} (1992) 2753.}\ it was demonstrated that there was a
one--to--one map between this open--closed string theory (which may be
formulated as a one--hermitain matrix model) and the double scaled
one--unitary matrix model. This model naturally has the $sl(2,{\rm C})$ mKdV
hierarchy
at its
heart and the Painlev\'e II hierarchy as its string equations. The map used was
the well--known Miura map,  relating the fundamental Hamiltonian structures of
the mKdV and KdV systems. Using the relation between the two systems and some
results from ref.\ref\nappi{T.Hollowood, L.Miramontes, C.Nappi and A.
Pasquinucci, Nucl. Phys. {\bf B373}, (1992) 247.}, the Virasoro constraints for
the open--closed string theory were derived.

The purpose of this paper is to fully elucidate the  structure of the $c<1$
open-closed string theories. They are shown to be intimately related to the
pure closed string theories and share and extend the underlying integrable
structure. The relation between the two families of theories is embodied in a
simple transformation between them, at the level of the integrable hierarchy.

The structure of the  $(2m-1,2)$ open--closed string theory, first studied to
all orders in ref.\kostovii, is
further
uncovered in section 2. After fixing a constant in the Virasoro constraints
derived  in
ref.\unitary, the structure of the constraints are understood in terms of the
underlying $Z_2$--twisted boson, and the $\tau$--function of the theory is
realised as the insertion of a vertex operator in the coherent state basis of
the boson.
The relation between the open--closed and closed string $\tau$--functions is
derived by
using a simple transformation of the background fields of the closed string
theory. The loop equations and pointlike operator recursion relations are
obtained from the Virasoro constraints and compared to the familiar ones for
the closed string theory. The model expected to be a $m=1$ topological
open--closed string theory is examined.
In particular the open--closed string puncture   equation is considered,
showing directly how the  puncture operators of the theories are simply related
to each other.
In section 3, the most straightforward generalisation of the model of section 2
is presented. This is the   open--closed string in the $(p,q)$ conformal
minimal model backgrounds whose non--Liouville
embedding
dimension naively consists of $q-1$ discrete points.
  The boundary conditions on this open--closed string are believed to be
trivial
punctual (Dirichlet) boundary conditions where each open--closed string loop is
embedded entirely
at  one of the `spacetime' points. The $W^{(q)}$ constraints on the
$\tau$--function of this theory are derived and the loop equations and operator
content are studied.
The example of the  open--closed string in the $(*,3)$ backgrounds is worked
out in some detail in section~4.    It is noted that the problem of correctly
identifying all of the relevant boundary operators expected from Liouville and
matrix model theory considerations is  present, as it is in the closed string
theory, and therefore the question\ref\bound{E.Martinec, G.Moore and N.Seiberg,
\pl\ {\bf B263} (1991) 190.}\ of whether the full physics of strings
propagating in $(p,q)$ conformal minimal model backgrounds is captured by the
integrable structures described here remains pertinent.  The paper ends with
some concluding remarks in section~5.

\section{The $(2m-1,2)$ open--closed string theory}
\subsection{Review}
It is not appropriate to review all of the details of the matrix model
derivations of the $(2m-1,2)$ open--closed string theory. These may be found in
ref.\kostovii. It is necessary however to unpack the notation  and  establish
the conventions to be used throughout this paper, and so a brief review of the
string equations and associated structures follows.

The string equation derived by Kostov\kostovii\ is equation
\openone. The object $\R$ is defined in equation \defr\ in terms of the
Gel'fand--Dikii differential polynomials. These are polynomials in $u$ and its
$x$--derivatives and are related via the recursion relation:
\eqn\recur{\R^{'}_{k+1}={1\over 4}\R^{'''}_k-u\R^{'}_k-{1\over 2}\R.}
They are fixed by determining the constant $\R_0$, and the requirement that
$\Rk\to 0$ as $u\to 0$. The first few are
\eqn\polly{\R_0={1\over2};\hskip 1cm\R_1=-{1\over4}u;\hskip
1cm\R_2={1\over16}(3u^2-u^{''}).}
Upon examination of the lowest member of the KdV flows \kdvflow\ the relation
between $x$ and $t_0$ is seen to be $x=-t_0/4$.
The diagonal part of the resolvent, defined as
\eqn\res{\Rs(x,\r)=<x|{1\over{-\nu^2\partial^2_z+u(z)+\r}}|x>}
has an expansion\Gelfand
\eqn\resexp{\Rs(x,\r)=\sum_{k=0}^\infty{\Rk[u]\over\r^{k+\hf}}}
and satisfies equation \geldik.
To study the $m$th model of the $(2m-1,2)$ series the $\tk$ in equation \defr\
are all set to zero except $t_0=-4x$ and $t_m$, which is set to a value which
cancels the coefficient of $u^m$ in $\Rm[u]$ to $1$. (This choice is equivalent
to a non--physical  rescaling of both $\nu$ and $x$.)
For example the string equation of the $(3,2)$ model is
\eqn\pgrav{-{1\over3}u^{''}+u^2-2\nu\Gamma\Rs(x,\r)=x.}  In the $\Gamma\to0$
limit this equation reduces to the Painlev\'e~I equation which reproduces the
genus expansion of $u$ in the $x\to+\infty$ limit. (Here $x$ is the
cosmological constant.) In order to derive the world sheet expansion for the
open--closed string theory using equations \geldik\ and \pgrav, an efficient
way to
proceed is to solve \pgrav\ for $\Rs$ and substitute it into \geldik, yielding:
\eqn\simply{(u+\r)\D^2-{1\over2}\D\D^{''}+{1\over4}(\D^{'})^2=\nu^2\Gamma^2,}
where $\D\equiv-{1\over3}u^{''}+u^2-x$. Simply expanding \simply\ for
$x\to+\infty$ yields the perturbation theory:
\eqn\pert{u=x^{1/2}-{1\over2}{\nu\Gamma\over x^{3/4}}-{1\over24}{\nu^2\over
x^2}+\cdots}
where the terms are the contributions from the sphere, disc and torus,
respectively. In general the asymptotic expansion takes the form\unitary
\eqn\asymp{u=x^{1/m}\sum_{g,h=0}^\infty A_{gh}{\nu^{2g+h}\Gamma^h\over
x^{(2+{1/m})(g+{h/2})}}}
where the indices $g$ and $h$ respectively represent the number of handles and
holes on the surface. The $A_{gh}$ are uniquely determined once the sphere
coefficient $A_{00}$ and the sign of the disc coefficient $A_{01}$ have been
fixed, using matrix model perturbation theory.

The function $u$ is related to the free energy $F$ of the model as follows:
\eqn\ufree{u(z)=\nu^2{\partial^2F\over\partial x^2},}
and is thus the connected two--point function of the puncture operator ${\cal
P}\equiv4{\cal O}_0$. The partition function $Z=\e{-F}$ is the square of the
$\tau$--function $\TO$ of the theory which is subject to the
constraints\unitary:
\eqn\consi{{\tilde L}_n\cdot\TO=0;\hskip 3cm n\ge-1}
where
\eqn\consii{{\tilde L}_n\equiv
L_n-(1+n){\Gamma^2\over4}\r^n-\r^{n+1}{\partial\over\partial\r}}
and\shoe{$^\dagger$}{Note here that in ref.\unitary\ the eigenvalue of the
$L_0$ constraint when acting on  $\TO$ was denoted $\mu$. Although it was shown
to be related to the open string coupling, the precise relation was not
determined there. It will be shown presently that $\mu=\Gamma^2/4$.}
\eqn\consiii{\eqalign{
L_{-1}&\equiv\sum_{k=1}^\infty(k+{1\over 2})t_k{\partial\over\partial
t_{k-1}}+{1\over4\nu^2}x^2\cr
L_0&\equiv\sum_{k=0}^\infty(k+{1\over 2})t_k{\partial\over\partial
t_k}+{1\over 16}\cr
L_n&\equiv\sum_{k=0}^\infty(k+{1\over
2})t_k{\partial\over\partial t_{k+n}}
+{4\nu^2}\sum_{k=1}^n{\partial^2\over\partial t_{k-1}\partial t_{n-k}}\hskip
1cm n\ge 1
.}}

The constraints for the closed string theory ($L_n\cdot\TC=0,\,\,n\ge-1$) are
recovered by setting $\Gamma=0$, and $\partial_\r\TO=0$ and then $\TO\to\TC$.
As first pointed out in refs.\ref\dijkone{R. Dijkgraaf, E.Verlinde and
H.Verlinde, Nucl. Phys. {\bf B348} (1991) 435.} and  \ref\kawai{M.Fukuma, H
Kawai and R.Nakayama,  Int. Jour. Mod. Phys. {\bf A6} (1991) 1385.}, the form
of the Virasoro generators \consiii\ are those of  a
$Z_2$--twisted boson whose mode expansion is:
\eqn\twist{\partial\phi(z)=\sum_{n=-\infty}^\infty\alpha_{n+\hf}z^{-n-3/2}}
where
\eqn\create{\alpha_{n+\hf}\equiv{2\sqrt2\nu}{\partial\over\partial t_n}\hskip
1cm{\rm and}\hskip 1cm\alpha_{-n-\hf}={1\over2\sqrt2\nu}(n+\hf)t_n.}
The energy--momentum tensor for the twisted boson is
\eqn\tensor{T(z)={1\over2}:\partial\phi(z)\partial\phi(z):+{1\over16z^2}
\hskip0.5cm=
\sum_{n=-\infty}^\infty L_nz^{-n-2}.}
Using \twist\ and \create\ in \tensor, the form \consiii\ is recovered for the
Virasoro generators.

The function $\TC$ defines a state $|\Omega_{\tilde t}>$ in the  Fock space of
the twisted boson\dijkone\ref\moore{G.Moore, Comm. Math. Phys. {\bf 133} (1990)
261.}\ and is defined as\shoe{$^\dagger$}{It is assumed here that a finite
number of the couplings $\tk$ are non--zero. Otherwise some of these
expressions are ill--defined\dijkone.}\eqn\closed{\TC=<t|\Omega_{\tilde t}>}
where $<t|\equiv<0|\exp{\{(2\sqrt2\nu)^{-1}\sum_{n=0}^\infty
t_n\alpha_{n+\hf}\}}$ defines the coherent state, upon which the annihilation
and creation operators act as derivatives and by multiplication respectively.
This twisted boson framework may be regarded as underpinning the structure of
the constraints of the closed string theory, together with the loop equations
and operator recursion relations which may be derived from them.

\subsection{{}From closed to open--closed strings}
The first observation to make about the open--closed string theory constraints
\consi\
is that they also admit the twisted boson description. Indeed, examination of
the form of these `modified' Virasoro operators in \consii\ suggest that
equations \consi\  are expressing the Ward identities of the Virasoro
generators in a conformal field theory\shoe{$^{\dagger\dagger}$}{The author
thanks Erik Verlinde for suggesting this interpretation.}. These may be derived
from the familiar operator product expansion of $T(z)$ with a conformal field
$\Phi_h(w)$ of weight $h$ as:
\eqn\ward{L_n\Phi_h(w)=(1+n)hw^n\Phi_h(w)+w^{1+n}\partial_w\Phi_h(w).}Therefore
the constraints \consi\ are suggestive of the presence of a conformal field in
the twisted sector of weight $\Gamma^2/4$. This would be most easily realised
as a vertex operator\shoe{$^{\dagger\dagger\dagger}$}{Some care must be
exercised here. The vertex operators of the form displayed for arbitrary $\r$
in the complex plane are not single valued as one encircles the origin, due to
the twist in $\phi(z)$.} of the form
$V_\Gamma(\r)=:\e{-{\Gamma\over\sqrt2}\phi(\r)}:$ This is the first hint of the
simple relationship between $\TO$ and $\TC$, the precise nature of which will
now be derived.

\def\ttk{\tilde t_k}
\def\tu{\tilde u}
\def\TT{\tilde\tau}
\def\tx{\tilde x}
Starting with the closed string theory with variables $\tk$ and functions
$u(\tk)$ and $\TC$, performing the change of variables:
\eqn\cov{\eqalign{t_k&\to\ttk+{2\Gamma\nu\over(k+\hf)}\r^{-(k+\hf)}\cr
x&\to\tx-\Gamma\nu\r^{-\hf}}} under which
\eqn\tild{u(t_k)\to \tu(\ttk,\Gamma,\r)\hskip1cm{\rm
and}\hskip1cm\TC(\tk)\to\TT(\ttk,\Gamma,\r),}
yields the string equation \eqn\opentwo{\R(\tu,\ttk)+2\Gamma\nu\Rs(x,\r)=0.}

Turning to the Virasoro constraints ($L_n(\tk)\cdot\TC=0,\,\,n\ge-1$) this
change of variables produces
\eqn\consiv{\eqalign{
&\sum_{k=1}^\infty(k+{1\over 2})\ttk{\partial\TT\over\partial
{\tilde t}_{k-1}}+2\Gamma\nu\sum_{k=1}^\infty\r^{-(k+\hf)}
{\partial\TT\over\partial{\tilde t}_{k-1}}
+\left({\tx^2\over4\nu^2}
-{\Gamma\over2\nu}\r^{-\hf}\tx+{1\over4}\Gamma^2\r^{-1}
\right)\TT=0\cr
&\sum_{k=0}^\infty(k+{1\over 2})\ttk{\partial\TT\over\partial\ttk}
+2\Gamma\nu\sum_{k=0}^\infty\r^{-(k+\hf)}
{\partial\TT\over\partial\ttk}
+{1\over 16}\TT=0\cr
&\sum_{k=0}^\infty(k+{1\over
2})\ttk{\partial\TT\over\partial {\tilde t}_{k+n}}
+2\Gamma\nu\sum_{k=0}^\infty\r^{-(k+\hf)}
{\partial\TT\over\partial{\tilde t}_{k+n}}
+{4\nu^2}\sum_{k=1}^n{\partial^2\TT\over\partial {\tilde t}_{k-1}\partial
{\tilde t}_{n-k}}=0\hskip 0.5cm n\ge 1
.}}

These may be drastically simplified by noting that
\eqn\deriv{{\partial\over\partial\r}\equiv
-2\Gamma\nu\sum_{k=0}^\infty\r^{-k-{3/2}}
{\partial\over\partial\ttk}}
and therefore
\eqn\trans{\eqalign{
&2\Gamma\nu\sum_{k=0}^\infty\r^{-(k+\hf)}
{\partial\TT\over\partial{\tilde t}_{k+n}}=\cr
&2\Gamma\nu\sum_{k=0}^\infty\r^{-(k-n+\hf)}
{\partial\TT\over\partial{\tilde t}_{k+n}}
-2\Gamma\nu\sum_{k=1}^n\r^{-(k-n-\hf)}
{\partial\TT\over\partial{\tilde t}_{k-1}}=\cr
&-\r^{n+1}{\partial\TT\over\partial\r}-2\Gamma\nu\sum_{k=1}^n\r^{-(k-n-\hf)}
{\partial\TT\over\partial{\tilde t}_{k-1}}
}}
Now writing $\TT=\e{g(\r,\ttk,\Gamma)}\TO$, a few more manipulations show that
\eqn\consv{{\tilde L}(\ttk)_n\cdot\TO=0,\hskip1cm n\ge-1,}
where the ${\tilde L}_n$ are given in equation \consii, for the unique form
\eqn\gee{g(\r,\ttk,\Gamma)={\Gamma^2\over4}\log\r
+{\Gamma\over4\nu}\sum_{k=0}^\infty\ttk\r^{k+\hf}.}
Note that the $L_0$ eigenvalue mentioned earlier is fixed as $\Gamma^2/4$ due
to the following
\eqn\due{\sum_{k=0}^\infty(k+\hf)\ttk{\partial g\over\partial\ttk}
-\r{\partial g\over\partial\r}=-{\Gamma^2\over4}.}
The relation between the functions $\TC$ and $\TO$ is now clear:
\eqn\relate{\TO=\e{-g(\tk,\r,\Gamma)}\TC\left(\tk\to\tk+
{2\Gamma\nu\over(k+\hf)}\r^{-(k+\hf)}\right),}
where now the tilde's above the open--closed string quantities have been
dropped, as
there can be no confusion.
Recalling equation \closed\ and using \relate\ yields
\eqn\vertex{\eqalign{\TO&=
<t|
\exp\left({-{\Gamma^2\over4}\log\r}\right)\times\cr
 &\exp\left({-{\Gamma\over\sqrt2}\sum_{k=0}^\infty{\alpha_{-k-\hf}
\over(k+\hf)}\r^{k+\hf}}\right)
 \exp\left({{\Gamma\over\sqrt2}\sum_{k=0}^\infty
{\alpha_{k+\hf}\over(k+\hf)}\r^{-(k+\hf)}}\right)
|\Omega_{\tilde t}>\cr
&=<t|:\e{-{\Gamma\over\sqrt2}\phi(\r)}:|\Omega_{\tilde t}>.
}}
So the open--closed string $\tau$--function is obtained from the closed string
$\tau$--function by the insertion of a vertex operator. This explains the form
of the constraints \consi, as anticipated at the beginning of this section.
(Notice that because $\r$ is restricted to be real by the world sheet
physics\shoe{$^\dagger$}{Recall that $\r$ is related to the boundary length
cosmological constant.}, the vertex operator is single valued.)

With the derivation of the open--closed string Virasoro constraints completed,
attention may now be turned to the study of some of the physics which may be
derived  from them, in the spirit of ref.\dijkone.  The recursion relations for
the pointlike operators in the theory together with the loop equations will be
derived next.

\subsection{Recursion relations}
The operators in the theory   are the $\Ok$, which couple to the $\tk$:
\eqn\correlate{<\O{k_1}\cdots\O{k_n}>=\left.
\nu^2{\partial^n\log Z\over\partial\t{k_1}\cdots\partial\t{k_n}}
\right|_{\tk=\ttk},} where $Z=\tau^2_{\rm open}$. Here $\ttk$ denotes some
fixed couplings to which the $\tk$ are set in the theory.
Recall from section~2 that the $m$th model in the $(2m-1,2)$ series is obtained
by setting all of the $\tk$'s to zero except $\t{m}=1/(\alpha_m(m+\hf))$ and
$t_0=-4x$, where the Gel'fand--Dikii polynomials start as
$\R[u]_m=\alpha_mu^m+\cdots$ To derive the recursion relations for some set of
operators in the $m$th theory, the Virasoro constraints \consi\ must be
expanded  in the couplings around these values. Furthermore, to obtain the
recursion relations on worldsheets of specific topology, the expansion
\def\S{\cal S}
 \eqn\expa{<\O{k_1}\cdots\O{k_n}>=\sum_{g,h=0}^\infty<\O{k_1}
\cdots\O{k_n}>_{g,h}
\nu^{2g+h}\Gamma^h} must also be made.
The derivation proceeds as follows. Set
\eqn\set{t_k={\delta_{mk}\over\alpha_m(m+\hf)}+
\epsilon_p\delta_{pk}-4x\delta_{k0},\hskip1cm m,p\neq0,\,\,p\in \S,}
where $\S$ represents some set of distinct operators and the $\epsilon_p$'s are
infinitesimal. Rearrange the constraints  to  act non--linearly on $\log\TO$
and  Taylor expand up to terms linear in the $\epsilon_p$'s. Then  all the
terms of order lower than $\prod_p\epsilon_p$ will be proportional to the
Virasoro constraints and may be set to zero leaving (after the world--sheet
expansion):
\def\cor#1#2#3{<#1\prod_{#2}#3>}
\def\OB{\O{B}}
\eqn\recursion{\eqalign{
&{1\over\alpha_m}\cor{\O{m+n}}{p\in\S}{\O{p}}_{g,h}+\sum_{p\in\S}(p+\hf)
\cor{\O{p+n}}{{r\in\S}\atop{r\ne p}}{\O{r}}_{g,h}\cr
-&4x\cor{\O{n}}{p\in\S}{\O{p}}_{g,h}
+4\sum_{k=1}^n
\left[\cor{\O{k-1}\O{n-k}}{p\in\S}{\O{p}}_{g-1,h}\right.\cr
+&\left.{1\over2}\sum_{{\S={\cal Q}\cup{\cal R}}\atop{g_1+g_2=g}}
\cor{\O{k-1}}{p\in\cal Q}{\O{p}}_{g_1,h}
\cor{\O{n-k}}{p\in\cal R}{\O{p}}_{g_2,h}
\right]\cr
-&\r^{n+1}<\O{\rho}\prod_{p\in\S}\O{p}>_{g,h}=0,
\hskip1cm n\ge-1,
}}
with the  exceptions on the sphere, torus and cylinder, respectively:
\eqn\except{\eqalign{&<\O{m-1}>_{0,0}=-{\alpha_mx^2\over2},\cr
&<\O{m}>_{1,0}=-{\alpha_m\over8},\hskip0.5cm {\rm and}\cr
&<\O{m+n}>_{0,1}={1\over2}(1+n)\alpha_m\r^n.}}
(Note that $\O{-1}=0$ in the above.) The operator $\O{\rho}$ is defined to be
that which couples to $\r$ in the following sense:
$<\O{\rho}>=\nu^2\partial_\r\log Z$.
Equation \recursion\ is the set of open--closed string recursion relations for
the
correlation functions of operators in the theory. Together with \except, they
are the generalisation of the closed string relations derived in ref.\dijkone\
and
have the same interpretation in terms of purely `contact' interactions. These
interactions take place when  the $\O{k}$'s coincide with each other, or in the
`factorization' situation, when the $\O{k}$ coincide with a node on the
surface. A node on the surface either pinches one of the handles, reducing $g$
by one, or divides the surface into two surfaces of total genus $g$. Notice
that there are no corresponding contact processes on the boundary, and so the
number of boundaries on the surfaces  are preserved. This is probably because
the $(2m-1,2)$ models
have only one distinct boundary boundary operator. This simplicity is expected
to disappear for more complicated theories. Also as before\dijkone, the
recursion relations may be used to determine the correlation functions of the
$\O{k}$ with $k\ge m-1$ in terms of the correlators of the $\O{k}$ with $k\le
m-2$. This latter set of operators may be identified with the (gravitationally
dressesd) primary fields of the $(2m-1,2)$ conformal field theory.

\subsection{A topological open--closed string and the boundary operator}
Setting $m=1$ in the closed string theory  yields topological gravity where the
surfaces in the underlying matrix model are not critical and are thus of zero
area in the continuum limit. The partition function of the theory is trivial as
can be seen by using the simple (exact) solution to the closed string equation,
$u=x$, to calculate the free energy via equation \ufree. After discarding
analytic terms in $x$ the result is $F=0$. The physical content of this
topological theory is studied at the level of correlation function recursion
relations.

 The
open-closed string generalisation of this theory is the $m=1$ model  of the
present framework. The   effect of adding surfaces with  boundaries  to the
theory is seen directly upon studying the string equation $u=x+2\Gamma\nu\Rs$,
together with the resolvent equation \geldik, which yields an expansion for the
free energy in which only surfaces with boundaries contribute:
\eqn\freemone{
F=\sum_{h>0,g\ge0}f_{gh}\left({\mu^{3/2}\over\nu}\right)^{2-2g-h}\Gamma^h
.}
Here $\mu=x+\rho$ is identified with the boundary cosmological constant (see
below) and the topological expansion is in a dimensionless combination of the
couplings in the theory\footnote{$^\dagger$}{The relative scaling dimensions of
all the couplings may be easiy determined from the matrix model or from the
string equation and flow equations.}.

The relation derived from \recursion\ for $n=-1$
is called the
`puncture equation' and is:
\eqn\puncture {
\cor{\OB}{p\in\S}{\O{p}}_{g,h}=
\sum_{p\in\S}(p+\hf)\cor{\O{p-1}}{{r\in\S}\atop{r\ne p}}{\O{r}}_{g,h}.}  %
%
%
%

Here the identifications $\OB=4\O{0}+\O{\rho}$ and  $\mu=x+\r$ of the boundary
operator and cosmological constant  respectively,  have been made. In this open
string theory therefore, the usual puncture equation is modified by  the
additive redefinition of  the puncture/boundary--length operator.

In the $m$th
model a similar equation to \puncture\ may be derived from the $n=-1$ part of
the recursion relation. {}From this it may be simply derived that the closed
string boundary operator\bound\ $\O{m-1}$ is modified to give the following
expression for the boundary operator:
\eqn\boundary{\OB=\O{\r}-{1\over\alpha_m}\O{m-1}.}
Using the puncture equation, the Ward identity of the boundary operator may be
easily verified be using the following expansion\ref\macro{T.Banks, M.Douglas,
N.Seiberg and S.Shenker, \pl {\bf B238} (1990) 279.}  of the macroscopic loop:
\eqn\exploop{w(\ell)=
\sum_{k=0}^\infty{\ell^{k+\hf}\over\Gamma(k+{3\over2})}\O{k},} yielding
\eqn\Wardii{<\OB w(\ell_1)\cdots
w(\ell_n)>=\left(\sum_{i=1}^n\ell_i\right)<w(\ell_1)\cdots w(\ell_n)>.}

\subsection{Loop equations}
Following ref.\dijkone\ it is convenient to introduce a source for the loops
which
shall be denoted $J(\ell)$:
\eqn\source{<w(\ell_{1})\cdots w(\ell_n)>=2\nu^2{\delta^n\log\TO\over\delta
J(\ell_1)\cdots\delta J(\ell_n)}.} Here the expression for the correlator is
interpreted as calculated in the presence of $J(\ell)$, and setting $J(\ell)=0$
returns one to the $m$th model. The Laplace transforms of these objects are:
\eqn\lapla{\eqalign{
w(z)&=\int_0^\infty d\ell\e{-\ell z}w(\ell)=\sum_{k=0}^\infty
z^{-k-{3\over2}}\O{k}\cr
J(z)&=\sum_{k=0}^\infty t_k z^{k+{1\over2}.}
}}
In ref.\dijkone\ the Virasoro constraints of the closed string theory arose as
the coefficients of the Laurent expansion of the Laplace transformed loop
equation ${\cal L}_{\rm closed}(z)=0$. Here, starting with the Virasoro
constraints for the open--closed string theory and interpreting them  as the
Laurent
coefficients of the open--closed string loop equations,
\eqn\laurent{
{\cal L}_{\rm open}(z)=2\nu^2\sum_{n=-1}^\infty
\left({{\tilde L}_n\cdot\TO\over\TO}\right)z^{-n-2}=0,
}
the following loop equation is derived:
\eqn\loop{\eqalign{
\int_0^\infty\!\!& d\ell^{'}\ell^{'}J(\ell)<w(\ell+\ell^{'})>
+{1\over8}\nu^2\ell+{1\over4}x^2
-\e{\r\ell}\left\{
{1\over4}\nu^2\Gamma^2
+<\O{\r}>\right\}\cr
+\int_0^\ell\!\!& d\ell^{'}
\left[4<w(\ell^{'})w(\ell-\ell^{'})>
+2<w(\ell^{'})><w(\ell-\ell^{'})>\right]=0.
}}
Here the modifications to the usual form of the closed string loop equation are
embodied in  an additional term arising from the cylinder\shoe{$^\dagger$}{This
new cylinder term joins the sphere and torus terms as the non-universal terms
arising from surfaces which admit global conformal transformations\dijkone.}
 together with an insertion of the operator $\O{\r}$.  Both terms have an
explicit exponential dependence on the loop length.

\section{The  open--closed string and $(p,q)$ minimal matter}
The discussion now turns to  open--closed strings in the   $(p,q)$ conformal
minimal model backgrounds. The aim of this section will be to use the insight
gained in
section~2 about the organisation of the $(2m-1,2)$ open--closed string to
derive
the
results here. This will have the effect of bypassing the matrix model route
altogether, remaining firmly within the integrable hierarchy framework. It is
not
entirely clear whether all of the physics of the related  $(p,q)$ conformal
minimal model backgrounds will be captured in this way; It has been
demonstrated  previously\bound\ that the  identification of the full spectrum
of boundary operators which is present in the matrix model and in the Liouville
approach for the gravitating $(p,q)$ minimal model is problematic for the
$sl(q,{\rm C})$ KdV approach. These difficulties were also encountered in
ref.\ref\pqmodels{C.Johnson, T.Morris and B.Spence, Nucl. Phys. {\bf B384}
(1992) 381.}\ where a stable non--perturbative definition of the closed string
in $(p,q)$ backgrounds was proposed, using the $sl(q,{\rm C})$ KdV as the
underlying
principle. There for example,  as in ref.\bound, only the boundary
magnetisation operator of the closed string in the Ising model background could
be identified. There appeared to be no structures left in the $ sl(3,{\rm C})$
flows
(and the associated $W^{(3)}$ constraints) within which to incorporate the
boundary length operator. This operator may be explicitly constructed in the
matrix model and Liouville approaches. This leads to the expectation that the
present  formulation of the  closed string in the $(p,q)$ conformal minimal
model backgrounds  based on the  integrable structure of the $sl(q,{\rm C})$
KdV
heirarchy captures only a subset of the boundary operators. Whether the missing
operators may be incorporated into a larger and/or complementary integrable
structure is unclear.
The open--closed string generalisation of the closed string just discussed
which will
be presented here will suffer from the same shortcomings (with respect to
boundary operators of the associated minimal model) and will likely be a small
subset of the family of open--closed strings which might be constructed with
these
backgrounds.

\subsection{Review}
The $sl(q,{\rm C})$ KdV hierarchies arise as the Hamiltonian flows of the $q$th
order
differential operator $Q=d^q+\sum_{i=2}^{q}\alpha_i\{u_i,d^{q-i}\}$. The symbol
``$d$'' denotes $\partial_x$ and the $u_i$ are functions of $x$. The $\alpha_i$
are constants. The study of the integrable deformations of $Q$ is most easily
formulated in terms of the  ``Lax pair'' where the Hamiltonian flows flows are
written in terms of fractional powers\ref\geldikii{I.M.Gel'fand and L.A.Dikii,
Funct. Anal.  Appl. {\bf 10} (1976) 259\semi I.M.Gel'fand and L.A.Dikii, Funct.
Anal.  Appl. {\bf 11} (1977) 93.}\ of $Q$
\eqn\hamflow{{\partial Q\over\partial t_r}=[Q_+^{r/q},Q].}
The non--trivial flows occur when $r\ne0\,\, {\rm mod}\,\, q$, i.e. when
$r=qk+l$, $k=0,1,\ldots,\infty$ and $l=1,2,\ldots,q-1$. The $t_r$ will be
labelled $\tlk$ to reflect this. The form of the flows analogous to \kdvflow\
is
(no sum on $i$): \eqn\qkdvflow{\alpha_i{\partial u_i\over\partial\tlk}={\cal
D}_1^{ij}\R^j_{l,k+1}={\cal D}_2^{ij}\R^j_{l,k}
\hskip 3cm i,j=2,3,\ldots,q.}
The objects $\Done$ and $\Dtwo$ are a set of differential operators reflecting
the bi--Hamiltonian structure which is expressed via the Poisson brackets:
\eqn\fund{\{u_i(x),u_j(y)\}_{1,2}={\cal D}_{1,2}^{ij}(x)\delta(x-y).}
It is well known (see, for example \ref\mattieu{P. Mathieu, \pl {\bf B208},
(1988) 101.}) that $\Dtwo$ has the structure of the classical W--algebra based
on $sl(q,{\rm C})$. This is the Hamiltonian system where the infinite set of
Hamiltonians is constructed out of traces of fractional powers of $Q$. The
 $\R^j_{l,k}$ are differential polynomials in the $u_i$ and their derivatives.
They are the generalisation of the Gel'fand--Dikii differential polynomials
encountered in the $q=2$ case in section~2. They are related to the
``jet--coefficients'' of the resolvent of $Q$ in the following way: The objects
defined by the expansion
\eqn\expii{\Rs^i=\bigsum\Rilk\r^{-(k+{l\over3})}}
are (a linear combination of) the ``$(q-i)$--jets'' of the resolvent
$\Rs(x,\r)$ which is defined by the relation
\eqn\res{\Rs(x,\r)\cdot(Q+\r)=1.}
For more details of this the reader should refer to the  work of Gel'fand and
Dikii\geldikii.  The $\Rs^i$ are shown there to satisfy an equation which
generalises equation~\geldik. It is easy to show that for the linear
combination of the ``jets'' of $\Rs(x,\r)$ which is used
here\shoe{$^\dagger$}{This linear combination is essentially defined by the
parametrisation used for $Q$ in terms of the $u_i$, together with equation
\qkdvflow.}, this  equation may be  shown to be:
\eqn\geldikiii{(\Dtwo+\r\Done)\Rs^j=0.}
 (Here an overall normalisation for the $\Dtwo$ will be chosen by setting
$\D^{i2}_2=u_id+{1\over i}u_i^{'}$.)
It is straightforward to demonstrate\pqmodels\ that by multiplying \geldikiii\
by $\Rs^i$ the  result is a total derivative. This is a consequence of the fact
that the infinite set of Hamiltonians of the system are in involution.
Integrating this equation once with respect to $x$ yields a single differential
equation for the $\Rs^i$'s which generalises equation \geldik, and deserves to
be called the `generalised Gel'fand--Dikii eqnation'. The constant of
integration is fixed by a normalisation  of the $\Rs^i$'s.

Using the expansion \expii\ in this equation yields a recursion relation for
the $\Rilk$ which is:
\eqn\recur{{\cal D}_1^{ij}\R^j_{l,k+1}={\cal D}_2^{ij}\R^j_{l,k}.}
Requiring them to vanish at $u_i=0$ fixes them uniquely, up to
the choice of normalisations ${\cal R}^i_{l,0}\propto q \delta^{i-1}_l$. This
normalisation fixes the constant of integration in the generalised
Gel'fand--Dikii equation to be $q^2$, as can be seen by substituting the
asymptotic expansion  \expii\ and taking the leading term in the $\r\to\infty$
and $z\to\infty$ limits. (This term always arises from the term
$(u_q+\r)\Rs^2\Rs^q$ in the equation.)

\nref\Ising{
V.A.Kazakov, \pl\ {\bf A119} (1986) 140\semi
D.V.Boulatov and V.A.Kazakov, \pl\ {\bf B186} (1987) 379\semi
E. Bre\'zin, M. Douglas, V. Kazakov and S. H. Shenker, Phys. Lett. {\bf B237}
(1990) 43\semi
D.J.Gross and A.A.Migdal, Phys.Rev.Lett {\bf 64} (1990) 717\semi
C.Crnkovi\'{c}, P.Ginsparg and G.Moore, Phys.Lett. {\bf B237} (1990) 196.}

Turning to the string equations for the closed string theory, the following
structures arise. The string equations arise from the differential operator
realisation of the canonical commutation relations\douglas\ $[P,Q]=1$. $Q$ is
as stated
above and the most general form for $P$ is constructed out of a linear
combination of the differential operator parts of fractional powers of $Q$. A
little study\pqmodels\ shows that the (once $x$--differentiated) string
equations which result from this construction may always be written as:
\eqn\stringeq{\Done\R^j=0,}
where:
\eqn\wh{\eqalign{\R^i&\equiv\sum_{l=1}^{q-1}\sum_{k=0}^\infty(k+{l\over
q})\tlk\Rilk\cr
&=\bigsum(k+{l\over q})\tlk\Rilk+(i-1)t_{i-1,0}}}
In the above the normalisation $\R_{l,0}^j=q\delta^{j-1}_l$ has been used. It
may also be shown that \stringeq\ is always a total derivative and may be once
integrated with respect to $x$ to give the string equation. The constants of
integration may always be absorbed into the $t_{l,0}$ which play special roles:
  For example the string in the   $(4,3)$ background has  $x\propto t_{1,0}$ as
the cosmological constant and $\B\propto t_{2,0}$ as the magnetic field. This
model is closely related to the string in a critical--Ising model background
(also denoted $(4,3)$ as it is a conformal minimal model). In fact the string
equations for it were first derived  by considering a matrix model realisation
of the gravitating Ising model\Ising. The string equation for the $p$th model
(denoted $(p,q)$)  is extracted from \stringeq\ by setting
\eqn\set{\tlk=-{\delta_{l,r}\delta_{k,s}\over(s+{r\over q})}+\sum_{r=1}^{q-1}
{q\over r}\delta_{l,r}\delta_{k,0}\tlk} where $p=qs+r$.

As an example of the above consider the $ sl(3,{\rm C})$ KdV system, better
known as
the Boussinesq heirarchy. The basic differential operator will be parametrised
as $Q=d^3+(3/4)\{u_2,d\}+u_3$ which results in the following flows for $u_2$
and $u_3$:
\eqn\qflowtwo{\alpha_{i}
{\partial u_i\over\partial\tlk}=\Done\R^j_{l,k+1}\equiv\Dtwo\Rjlk \hskip
3cm i,j=2,3.}
Here $\alpha_{2}=3/2,\alpha_{3}=1$, and there is no sum on $i$. Also\pqmodels:
$$\eqalign{
&\D^{22}_2={2\over 3}d^3+{1\over 2}u_2^{'}+u_2d\cr
&\D^{23}_2=u_3d+{2\over 3}u_3^{'}\cr
&\D^{32}_2=u_3d+{1\over 3}u_3^{'}\cr
&\D^{33}_2=-{1\over 18}d^5-{5\over 12}u_2d^3-{5\over 8}u_2^{'}d^2
+(-{1\over 2}u_2^2-{3\over 8}u_2^{''})d+(-{1\over 2}u_2u_2^{'}-{1\over
12}u_2^{'''})
\cr
}$$
and $\D^{22}_1=\D^{33}_1=0;\hskip 1cm\D^{23}_1=\D^{32}_1=d$.

A few of the differential polynomials are\pqmodels:
$$\eqalign{
&\R^2_{1,0}=3;\hskip 2cm\R^3_{1,0}=0;\cr
&\R^2_{2,0}=0;\hskip 2cm\R^3_{2,0}=3;\cr
&\R^2_{1,1}=u_3;\hskip 1.6cm\R^3_{1,1}={3\over 2}u_2;\cr
&\R^2_{2,1}=-{1\over 4}(u^{''}_2+3u_2^2);\hskip 1.6cm\R^3_{2,1}=2u_3;\cr
&\R^2_{1,2}=-{1\over 12}u^{''''}_2-{3\over 4}u_2u^{''}_2-{3\over 2}(u_2^{'})^2
-{1\over 2}u^3_2+{4\over 3}u^2_3;\hskip 0.5cm\R^3_{1,2}={2\over
3}(u^{''}_3+3u_3u_2);\cr
}$$
The string equations are:
\eqn\streq{\eqalign{
&\R^2=\sum_{{k=0}\atop{l=1,2}}^\infty(k+{l\over3})\Rlk^2-x=0\hskip1cm{\rm
and}\cr
&\R^3=\sum_{{k=0}\atop{l=1,2}}^\infty(k+{l\over3})\Rlk^3-\B=0.
}} Here $x=-t_{1,0}$ and $\B=-2t_{2,0}$.

The objects $\Rs^2$ and $\Rs^3$ satisfy the following (generalised
Gel'fand--Dikii) equation:
\eqn\strng{
\eqalign{
&{1\over 2}u_2\Rs_2^2+{2\over 3}\Rs_2\Rs^{''}_2-{1\over 3}(\Rs^{'}_2)^2
+u_3\Rs_2\Rs_3
-{1\over 18}\left(\Rs_3\Rs^{(4)}_3-\Rs^{'}_3\Rs^{'''}_3-{1\over
2}(\Rs^{''}_3)^2\right) \cr
&-{5\over 12}\left(u_2\Rs_3\Rs^{''}_3-{1\over 2} u_2(\Rs^{'}_3)^2+
{1\over 2} u_2^{'}\Rs_3\Rs^{'}_3\right)-{1\over
12}\left(3u_2^2+u_2^{''}\right)\Rs^2_3=9 \cr
}
}
(For convenience of notation  the superscripts on the $\Rs^i$'s  have been
exchanged for subscripts.)

The generalisation of the Virasoro constraints on the $(p,q)$ models is
phrased most naturally in terms of twisted bosons. These are the
$W^{(q)}$--algebra constraints. The underlying structure of the
$W^{(q)}$--algebra constraints on the closed string $\tau$--function $\TC$ is
that of the coherent state realisation of these constraints on the Fock space
of $q-1$ free bosons with the $Z_q$--twisted boundary conditions:
\eqn\twist{\partial\phi_l(\e{2\pi i}z)=\e{{2\pi il\over q}}\phi_l(z).}
They have the mode expansion
\eqn\mode{\partial\phi_l(z)=
\sum_{k=-\infty}^{\infty}\alpha_{k+{l\over q}}z^{-(k+{l\over q}+1)},
} where
\eqn\alp{\alpha_{k+{l\over q}}=\nu{\del\over\del\tlk}\hskip0.5cm{\rm
and}\hskip0.5cm\alpha_{-k-{l\over q}}={1\over\nu}(k+{l\over q})\tlk.}
The energy--momentum tensor is
\eqn\emt{T(z)=\sum_{r=1}^{q-1}:{1\over2}\partial\phi_r\partial\phi_{q-r}:
+{q^2-1\over24qz^2}=\sum_{n=-\infty}^{\infty}L_nz^{-n-2}.} The $W^{(q)}$
operators in the constraints arise as the modes $W^{(r)}_n$ of the higher spin
fields $W^{(r)}$  ($r=3,\ldots,q$), constructed using for example,
current--algebras based on $sl(q,{\rm C})$. Denoting $T(z)$ as $W^{(2)}$ ($L_n$
as
$W^{(2)}_n$), the closed string constraints are
\eqn\wcons{W^{(r)}_n\cdot\TC=0,\hskip1cm r=2,\cdots,q,\,\,n\ge1-r,}
where $\TC$ defines a state $|\Omega_{\tilde t}>$ in the coherent state basis
via
\eqn\closed{\TC=<t|\Omega_{\tilde t}>.} Here
\eqn\teatime{
<t|=<0|\exp\left\{{1\over\nu}\bigsum\tlk\alpha_{k+{l\over q}}\right\}.}
Note that the functions $u_i$ are related to the $\tau$--function as follows
(no sum on $i$):
\eqn\rel{\alpha_iu_i=2\nu^2{\del^2\log\TC\over\del t_{i-1,0}\del t_{1,0}},}
and the operators in the theory arise as
\eqn\operate{<\O{l_1,k_1}\cdots\O{l_n,k_n}>
=2\nu^2{\del^n\log\TC\over\del t_{l_1,k_1}\cdots\del t_{l_n,k_n}}.}
More details of the constraints may be found in
refs.\dijkone\kawai\ref\goeree{J.Goeree, \nuke\ {\bf B358} (1991) 737.}.

\subsection{{}From closed to open--closed strings}
The starting point is to generalise the transformation on the $\tlk$ analogous
to equation \cov. This is simply\shoe{$^\dagger$}{The factor of 2 in the
transformation \cov\ is present in order to match the conventions of
ref.\unitary.}
\def\ttlk{{\tilde t}_{l,k}}
\eqn\cortwo{\tlk\to\ttlk+{\Gamma\nu\over(k+{l\over q})}\r^{(k+{l\over q})}}
and consequently
\eqn\conseq{
u_i(\tlk)\to{\tilde u}_i(\tlk,\Gamma,\r).
}
The string equation under these transformations become
\eqn\streqiv{\Done\left[\R^j({\tilde u})+\Gamma\nu\Rs^j(x,\r)\right]=0,}
which  may be integrated once to become the open--closed string generalisation
of
\opentwo, and must be supplemented by the generalised Gel'fand--Dikii equation
\geldikiii\ to get solutions.

The most straightforward method by which the open--closed string constraints
may be
derived is by returning to the twisted bosons and using the vertex operator
construction used in section~2. Then it is seen by direct calculation that  the
transformation \cortwo\ is the beginning of the process by which the
combination $\e{-\Gamma\sum_{l=1}^{q-1}\phi_l(\r)}$ is inserted into the
bra--ket~\closed\ to give the open--closed string $\tau$--function. Explicitly:
\eqn\open{
\TO=\e{-g(\tlk,\rho,\Gamma)}\TC\left(\tlk\to\tlk+{\Gamma\nu\over(k+{l\over
q})}\r^{(k+{l\over q})}\right)
,}
where
\eqn\geee{g(\rho,\tlk,\Gamma)={\Gamma^2\over2}\log\r+{\Gamma\over\nu}
\bigsum\tlk\r^{k+{l\over q}}.}
Consequently
\eqn\Vertex{\eqalign{\TO&=
<t|\exp\left({-{\Gamma^2\over2}\log\r}\right)\times\cr
&\exp\left({-{\Gamma}\bigsum{\alpha_{-k-{l\over q}}\over(k+{l\over
q})}\r^{k+{l\over q}}}\right)
\exp\left({{\Gamma}\bigsum{\alpha_{k+{l\over q}}\over(k+{l\over
q})}\r^{-(k+{l\over q})}}\right)
|\Omega_{\tilde t}>\cr
&=<t|V_\Gamma(\rho)|\Omega_{\tilde t}>,
}}
where
 \eqn\Vert{V_\Gamma(\r)=
:\exp\left\{{-{\Gamma}\sum_{l=1}^{q-1}\phi_l(\r)}\right\}:}
To work out the constraints on $\TO$ is simply a matter of calculating the
operator product expansion
\eqn\operate{W^{(r)}(z)V_\Gamma(w)=
N(r){\Gamma^r\over(z-w)^r}V_\Gamma(w)+\cdots\,\,r=2,\cdots,q}
where  $N(r)$ is a numerical constant. The other terms will in general be a
combination of $V_\Gamma(w)$, the $\phi_l(w)$ and their $w$--derivatives.
Performing the mode decomposition \eqn\decom{W^{(r)}_n=\oint
dz\,\,z^{n+r-1}W^{(r)}(z)}
yields the form
\eqn\decomp{[W_n^{(r)},V_\Gamma(w)]
\sim\Gamma^r[(n+1)(n+2)\cdots(n+r-1)]w^nV_\Gamma(w)+\cdots}
The $W^{(q)}$ constraints for the open--closed string theory are then
\eqn\opencon{{\widetilde W}^{(r)}_n\cdot\TO=0\hskip1cm
r=2,3,\cdots,q;\,\,\,\,n\ge1-r,}
where
\eqn\condef{{\widetilde W}^{(r)}_n=W^{(r)}_n-[W^{(r)}_n,V_\Gamma(\r)].}
These equations are the direct extension of equations \consi\ and \consii.
To study the loop equations of the theory, the constraints \opencon\ are used
as Laurent coefficients in the expansion of the Laplace transform of $q-1$ loop
equations:
\eqn\loopi{{\cal L}^{(r)}_{\rm open}(z)=2\nu^2\sum_{n=1-r}^\infty\left(
{{\widetilde W}^{(r)}_n\cdot\TO\over\TO}\right)z^{-n-r}=0}
and the Laplace transformed loop variables are extracted in analogy with
\exploop\ as follows\dijkone\goeree:
\eqn\well{\eqalign{
w_r(z)&=\sum_{k=0}^\infty z^{-k-{r\over q}-1}\O{r,k}\cr
J_r(z)&=\sum_{k=0}^\infty t_{r,k} z^{k+{r\over q}}.
}}
 Inverse Laplace transforming \loopi\ yields the final loop equations for the
loops
\eqn\loopii{w_r(\ell)=\sum_{k=0}^\infty{\ell^{k+{r\over q}}\over
\Gamma(k+{r\over q}+1)}\O{r,k}} and their sources $J_r(\ell)$.

\subsection{Recursion relations--A boundary operator}
The recursion relations for the correlators of the operators in the $(p,q)$th
model open--closed  string theory may be derived by expanding in the
couplings as
performed in section~2. The structure of the constraints is such that there
arise $q-1$ sets of recursion relations, the ${\widetilde W}^{(l-1)}$ sector of
the constraints (see equation \opencon)  resulting in the recursion relations
between correllators of the operators $\O{l,k}$. See ref\goeree\ for an
explicit example of this structure in the case of the $(1,3)$ closed string
model. The form of these constraints is in general highly complicated, although
they readily allow the  dressed primary fields of the underlying conformal
minimal model to be identified as the relevant operators in the
bulk\dijkone\goeree. Here the general relations will not be analysed any
further and attention will be turned instead to the recursion relation arising
from the $L_{-1}$ constraints. In particular, the unitary ($q+1,q$) models will
be studied. The unitary model of the   series is obtained by
setting
\eqn\setting{\tlk=-{q\delta_{l.1}\delta_{k,2}\over(2q+1)}+\sum_{r=1}^{q-1}
{q\over r}\delta_{l,r}\delta_{k,0}t_{r,0}.}
Now the $L_{-1}$ constraint for the $q$th model is of the form
\eqn\ellmone{\bigsum(k+{l\over q})\tlk{\del\TO\over\del t_{l,k-1}}
+\nu^{-2}f(\tlk)=0,}
where  $f(\tlk)$ is constructed out of products of the $\tlk$'s only.  It will
eventually become the spherical contribution to the non--universal parts of the
correlation functions. Such terms will not be of concern in what follows.
Expanding in the couplings about \setting\ yields:
\eqn\Bound{\cor{\OB}{r,s\in S}{\O{r,s}}=\sum_{r,s\in S}(s+{r\over
q})\cor{\O{r,s-1}}{{a,b\in S}\atop{a\ne r,b\ne s}}{\O{a,b}},} where
\eqn\Boundary{\OB=\O{1,1}-\O{\r}} is defined in analogy with \boundary.

In ref.\bound\ the operator $\O{1,1}$ was identified with a boundary operator
in the closed string unitary conformal minimal model backgrounds. Equation
\Boundary\ gives the modification in the open--closed string theories here.
Notice that
using the expansion \well\ a Ward identity identical in structure to \Wardii\
may be derived for $\OB$ here. Taking the Ising model $(4,3)$ as an example
this relation implies that $\OB$ measures the total length of the loops in the
correlator, regardless of which combination of the two loops $ w_1(\ell)$ or
$w_2(\ell)$ appears. This appears at first to be in contradiction of the
identification of $\O{1,1}$ as the $Z_2$--odd Ising--string boundary
magnetisation in ref.\bound. This is resolved if it is recalled from there that
the loop operators $w_+(\ell)$ and $w_-(\ell)$ given by
\eqn\isingloop{w_\pm(\ell)=\int_\mu^\infty<x|\e{\pm\ell Q}|x>\,dx}are to be
considered as the `correct' loop operators for the Ising--string arising from
the underlying matrix model. Under the $Z_2$ operation $Q\to-Q$, $u_2\to u_2$,
$u_3\to-u_3$ and $w_+(\ell)\to w_-(\ell)$. Considering a shift in $u_3$ results
in identifying $\O{1,1}$ as the boundary magnetisation\bound:
\eqn\magnet{<\O{1,1}w_\pm(\ell)>=\pm\ell<w_\pm(\ell)>.}
The loop operators $w_1(\ell)$ and $w_2(\ell)$ in this integrable formalism are
thus clearly not the Ising--string loops above. To derive a relation like
\magnet\ from the $L_{-1}$ constraint \ellmone\ would require an appropriate
expansion of $w_+(\ell)\to w_-(\ell)$ in the $\O{l,k}$ to replace \well. This
expansion would have to exhibit the $Z_2$ structure of the underlying Ising
model. Some of these issues are discussed in ref.\ref\mooretwo{G.Moore,
N.Seiberg and M.Staudacher, Nucl. Phys. {\bf B362} (1991) 665.} where more
understanding of the change of bases required to relate the Liouville, matrix
model and integrable hierarchy approaches to various string models is
presented.

It is tempting to speculate about the role of the constraints ${\widetilde
W}^{(r)}_{-r+1}$.
Regarding each constraint as the Ward identity of an operator, it would be
interesting to understand the  nature of the operators defined by these
constraints. For example in the case of $q=3$, ${\widetilde W}^{(3)}_{-2}$ has
precisely the  $Z_2$--even grading required for the boundary length operator in
the Ising--string\pqmodels. The Ward identity derived from this constraint does
not have such a simple structure however. Nevertheless perhaps the
understanding of how to define the correct Ising loop operators in the $(4,3)$
model would also reveal whether or not ${\widetilde W}^{(3)}_{-2}$ defines the
boundary length operator. This would go some way to understanding just how much
of the $(p,q)$ conformal minimal model is contained in this integrable
formulation.

\section{The $sl(3,{\rm C})$ structure of the $(*,3)$  models }
It is now time to turn to a specific example of the constructions just
described and study the structure of the constraints and  loop equations of the
 $(*,3)$ backgrounds. In the spirit of refs.\dijkone\ and \goeree\ the
subtle
matter of identifying the correct frame in which to discuss the loop variables
in the underlying minimal model is sidestepped in favour of continuing to study
the integrable structure. Therefore the loop equations derived
will be in terms of the objects $w_1(\ell)$ and $w_2(\ell)$.

\bigskip

\subsection{$W^{(3)}$ constraints}
The form of the spin 2 and spin 3 fields $W^{(2)}(z)$, ($T(z)$), and
$W^{(3)}(z)$ for the two $Z_3$ twisted bosons $\phi_1(z)$ and $\phi_2(z)$ are:
\eqn\spins{\eqalign{
T(z)&=:\partial\phi_1(z)\partial\phi_2(z):+{1\over9z^2}\cr
W^{(3)}(z)&=:(\partial\phi_1(z))^3+(\partial\phi_2(z))^3:
}} and calculation of the operator products and mode decompositions using the
ideas and formulae in section 3 yields:
 \eqn\constri{\eqalign{
{\tilde L}_n\cdot\TO&=0\hskip1cm n\ge-1;\cr
{\tilde W}_m\cdot\TO&=0\hskip1cm m\ge -2,
}}
where
\eqn\constrii{\eqalign{
{\tilde L}_n
=L_n&-{\Gamma^2\over2}(n+1)\r^n-\r^{n+1}{\partial\over\partial\r};\cr
{\tilde W}_m=W_m&-{\Gamma^3\over12}(n+1)(n+2)\r^n+{\Gamma^2\over4}(n+2)\r^{n+1}
{\partial\over\partial\r}+\cr
&\hskip1cm\left[{\Gamma^2\over4}(\partial^2\phi_1(\r)+\partial^2\phi_2(\r))
-{\Gamma\over4}((\partial\phi_1(\r))^2+(\partial\phi_2(\r))^2)\right]\r^{n+2}.
}} The decomposition of the  $L_n$ and $W_m$ into the modes of the bosons
yields
\def\some{\sum_{{k=0}\atop{l=1,2}}^\infty}
\def\diff#1{{\partial\over\partial#1}}
\eqn\decompi{\eqalign{ L_{-1}&=\sum_{{k=1}\atop{l=1,2}}^\infty
(k+{l\over3})\tlk{\partial\over\partial
t_{l,k-1}}+{2\over3\nu^2}t_{1,0}t_{2,0}\cr
L_0&=\some(k+{l\over3})\tlk\diff{\tlk}+{1\over9}\cr
 L_n&=\some(k+{l\over3})\tlk\diff{t_{l,k+n}}+
{1\over6}\nu^2\sum_{{k=1}\atop{l=1,2}}^n{\partial^2\over\partial
t_{l,k-1}\partial t_{3-l,n-k}}
}} and\goeree
\eqn\decompii{\eqalign{
 W^{(3)}_m=&\nu^{-3}\sum_{p+q+r=-3m}{p\over3}
{q\over3}{r\over3}t_pt_qt_r+\nu^{-1}\sum_{p+q-r=-3m}{p\over3}
{q\over3}t_pt_q\diff{t_r}\cr
+&{1\over3}\nu\sum_{-p+q+r=3m}{p\over3}t_p{\partial^2\over\partial t_q\partial
t_r}+
{1\over27}\nu^3\sum_{p+q+r=3m}{\partial^3\over\partial t_p\partial t_q\partial
t_r}
.}}In the last formula, the sums must be taken over $p,q,r\ne0$ mod 3.
The constraints \constri\ for the open--closed string theory still satisfy the
familiar
commutation relations  of the $W^{(3)}$ algebra and are
thus a consistent set of constraints.

\subsection{Loop equations}
The set of constraints \constri\ will yield a pair of basic loop equations for
the $(*,3)$ open--closed string theory which will be the closed string loop
equations\goeree\dijkone\ with additional terms, as encountered in section 2.
{}From the Virasoro sector the following equation is obtained (sum over $r$):
\eqn\loopp{\eqalign{
\int_0^\infty\!\!& d\ell^{'}\ell^{'}J_r(\ell)<w_r(\ell+\ell^{'})>
+{2\over9}\nu^2\ell+{4\over3}t_{1,0}t_{2,0}
-\e{\r\ell}\left\{
{1\over2}\nu^2\Gamma^2
+<\O{\r}>\right\}\cr
+\int_0^\ell\!\!& d\ell^{'}
\left[4<w_r(\ell^{'})w_{3-r}(\ell-\ell^{'})>
+2<w_r(\ell^{'})><w_{3-r}(\ell-\ell^{'})>\right]=0.
}}
and from the $W^{(3)}$ sector (sum over $r$):
\eqn\looppp{\eqalign{
\int_0^\infty\int_0^\infty\!\!d\ell^{'}d\ell^{''}&(\ell^{'}J_r(\ell^{'}))
(\ell^{''}J_r(\ell^{''}))<w_{3-r}(\ell+\ell^{'}+\ell^{''})>\cr
+&{16\over27}t_{2,0}^3+{8\over9}t_{1,0}^2t_{1,1}+{2\over27}t_{1,0}^3\ell+
 \e{\r\ell}\left({\Gamma^2\over4}\ell<\O{\r}>-
{\nu^2\Gamma^3\over12}\ell^2\right) \cr
+&\e{\r\ell}\left[
{\nu^2\Gamma^2\over2}\partial^2\phi_r(\r)-
{\nu^2\Gamma\over2}(\partial\phi(\r))^2\right]\cr
 +{1\over6}\int_0^\infty\int_0^{\ell+\ell^{'}}\!\!&d\ell^{'}d\ell^{''}
(\ell^{'}J_{3-r}(\ell^{'}))\times\cr
&\left[\nu^2<w_r(\ell^{''})w_r(\ell+\ell^{'}-\ell^{''})>
+{1\over2}<w_r(\ell^{''})><w_r(\ell+\ell^{'}-\ell^{''})>\right]\cr
 +{1\over27}\int_0^\infty\int_0^{\ell-\ell^{'}}\!\!&d\ell^{'}d\ell^{''}
\left[
\nu^4<w_r(\ell^{'})w_r(\ell^{''})w_r(\ell-\ell^{'}-\ell^{''})>\right.\cr
+&{3\over2}\nu^2<w_r(\ell^{'})><w_r(\ell^{''})w_r(\ell-\ell^{'}-\ell^{''})>\cr
+&\left.{1\over4}<w_r(\ell^{'})><w_r(\ell^{''})><w_r(\ell-\ell^{'}-\ell^{''})>
\right]=0.
}}
Equation \loopp\  which arises from the Virasoro sector of the constraints has
the same interpretation as in the previous, $(2m-1,2)$ case. It describes
how
the loops interact by splitting and joining\dijkone\ref\loopss{S.Wadia, Phys.
Rev. {\bf
D24} (1981) 970\semi A.Migdal, Phys. Rep. {\bf 102} (1983) 199\semi V. Kazakov,
Mod. Phys. Lett. {\bf A4} (1989) 2125\semi F. David, Mod. Phys. Lett. {\bf A5}
(1990) 1019.}.
 These processes are taking place on surfaces of arbitrary Euler number. There
is an additional non--universal term arising from the cylinder, which together
with the sphere and the torus, admits global conformal transformations. There
is also an insertion of the operator $\O{\r}$, as before. The closed string
part of
\looppp\ which arises from the W--sector of the constraints were first
presented in ref.\goeree. Equation \looppp\ describes the three--way splitting
and joining of loops on the world sheets, and is supplemented by a number of
terms, including an insertion of $\O{\r}$ and additional contributions from the
sphere, torus and cylinder. The precise interpretation of the terms involving
$\partial\phi_r(\rho)$ is not clear although recalling the relation between
$\del\phi_r(\r)$ and the loop variables:
\eqn\phloop{\nu\del\phi_r(\r)=<w_r(z)>+\partial J_{3-r}(z)} allows the loop
equation to be rewritten in terms of loop variables only. The presence of the
$\del\phi_r(\r)$ terms may suggest some non--trivial interactions at the
boundaries of the surfaces.

\section{Conclusions}
The  integrable structure of the $c<1$ open--closed string theories described
in this paper is
the most
straightforward generalisation of the structure of their more familiar  $c<1$
closed
string cousins. They
trivially reduce to the closed string in the $\Gamma\to0$ limit where $\Gamma$
is the open string coupling measured in units of the closed string coupling
$\nu$. They inherit many of the properties of the closed string. Chief of these
is the relation to a string in the $(p,q)$  conformal minimal model
backgrounds. This relation is incomplete because the task of identifying the
boundary operators (for $q>2$) is incomplete, even for the closed string
theories. For the
unitary models, the boundary operator
which may always be identified in the closed string theory, $\O{1,1}$, is
modified in the open--closed string theory, as expected. It is modified by
adding the
operator which couples to $\r$ in the open--closed string theory. $\r$ is the
spectral
parameter of the $sl(q,{\rm C})$ KdV hierarchy.
The open--closed string also inherits the embedding dimension of the closed
string
because of the simple relationship between  the two theories. The loops of the
closed string come in $q-1$ distinct `colours'. This degreee of freedom may be
naively interpreted as a discrete embedding dimension consisting of $q-1$
points. The boundary conditions on the open--closed string are
simply the trivial punctual (Dirichlet) ones, where each boundary on a world
sheet is
entirely of one of the $q-1$ colours.

 The organisational framework for both the open--closed and closed strings is
the
twisted boson language. The open--closed string partition function is obtained
from the
closed string one by inserting a vertex function constructed out of the twisted
bosons into the Fock--space realisation of the $\tau$--function. As this
procedure is `orthogonal' to the flows of the underlying integrable structure,
the resulting object is still a $\tau$--function. It is using this language
that enables the constraints on the partition function to be easily deduced.

Turning to the recursion relations for correlators, some related work on the
$(2m-1,2)$ open--closed string theory was presented in
ref.\ref\tanii{Y.Itoh and Y.Tanii, Phys. Lett. {\bf 289} (1992) 335.}. There,
the Dyson--Schwinger equations of the theory were derived by working directly
with the matrix model of Kostov described earlier. Recursion relations were
derived there for pointlike operators related to the $\O{k}$ discussed in
section~2. By defining a generating function $\tau$ for the insertion of the
operators, the recursion relations were able to be written as constraints on
$\tau$. It was also noted there that a shift of the sources of the operators,
identical to \cov, could be made which removed the open string variables. A
further change of variables enabled the constraints to be written as the
familiar (closed string) Virasoro constraints. This series of
re--identifications of the sources is precisely the reverse of the process
carried out in section~2.2 to relate the closed and open--closed string
$\tau$--functions. However the generating function $\tau$ identified in
ref.\tanii\ is
$\TO$ in equation \vertex\ with the exponential factor missing.  Taking this
into account their
original constraints may be rewritten in terms of $\TO$, and this yields the
open--closed string Virasoro
constraints \consi, which were first derived in ref.\unitary. Once the
constraints are written in that form,   the
underlying vertex operator structure is more evident, as revealed in this
paper.

It is a straightforward matter to construct such simple open--closed string
generalisations of closed strings based on integrable hierarchies other than
$sl(q,{\rm C})$  using the same techniques, although the continuum string
theory interpretation of such possible models is not at present known. The
essential relation between the
open--closed and closed string theory here is the transformation on the
background
fields of the closed string theory to obtain those of the open--closed string
theory.
This is given by equations \cov\ and \cortwo. In the integrable hierarchy, this
is
simply a transformation on the generalised  times of the evolution equations.
Of particular interest and value  would be the application of the techniques
and understanding developed in this paper to the problem of understanding the
nature of the $c=1$ open--closed string. For $c=1$,  some of the integrable
structure has been understood (see for example ref.\ref\dijk{R.Dijkgraaf,
G.Moore and R.Plesser, Nucl. Phys. {\bf B394} (1993) 356.}.). Notably though,
the $c=1$ open--closed string theory is not developed to the same level as the
closed string theory\footnote{$^\dagger$}{See for example,
refs.\ref\opensi{D.V.Boulatov, Int. Jour. Mod. Phys. {\bf A6} (1991) 79.}  to
\nref\opensii{Z.Yang, Phys. Lett. {\bf B257} (1991) 40.}
\nref\openiii{M. Bershadsky and D. Kutasov, Phys. Lett. {\bf B274} (1992) 331.}
\nref\opensiv{M. Bershadsky and D. Kutasov, Nucl. Phys. {\bf B382} (1992) 213.}
\ref\opensv{J.A. Minahan, University of Virginia preprint UVA--HET--92--01,
hepth 9204013.}.}. These issues are the subject of current research.

\vskip 2cm

{\bf Acknowledgements.}
The author is grateful to Simon Dalley, M{\aa}ns Henningson, Jeremy Schiff and
 Erik Verlinde for illuminating discussions. Many thanks also to Simon Dalley,
Chantal Morgan, Tim Morris and Samson Shatashvilli for comments on the
manuscript. This research was
financially supported by a Lindemann Fellowship.

\vskip 1cm

{\bf Note Added.}
After this work was submitted for publication, L.Houart\ref\houart{L.Houart,
Bruxelles preprint ULB--TH--02/93, hepth 9303157.}\ studied a two--hermitian
matrix model realisation of an open--closed string in the Ising model
background, with Dirichlet boundary conditions on the ends of the open strings.
The double scaling limit of this matrix model was shown to yield the string
equations derived in section three of this paper for the $(4,3)$ open-closed
string model. This supports the expectation mentioned earlier that the boundary
conditions on these open--closed strings are Dirichlet.

\listrefs
\bye